\documentclass[letterpaper, 10pt, conference]{ieeeconf}  

\IEEEoverridecommandlockouts                              

\usepackage{graphicx}
\usepackage{url}  
\usepackage{amsmath}
\usepackage{todonotes}

\DeclareMathOperator*{\argmax}{argmax}
\DeclareMathOperator*{\softmax}{softmax}

\begin{document}

\title{\LARGE \bf
	CNN-MoE based framework for classification of respiratory\\anomalies and lung disease detection
}

\author{Lam~Pham, 
	Huy~Phan, 
	Ramaswamy~Palaniappan,
	Alfred~Mertins,
	Ian~McLoughlin
	\thanks{L. Pham and R. Palaniappan are with the School of Computing, University of Kent, UK. H. Phan is with the School of Electronic Engineering and Computer Science, Queen Mary University of London, UK. A. Mertins is with the Institute for Signal Processing, University of L\"ubeck, Germany. I. McLoughlin is with Singapore Institute of Technology, Singapore.}%
}

\maketitle
\thispagestyle{empty}
\pagestyle{empty}

\begin{abstract}
This paper presents and explores a robust deep learning framework for auscultation analysis. This aims to classify anomalies in respiratory cycles and detect disease, from respiratory sound recordings.
The framework begins with front-end feature extraction that transforms input sound  into a spectrogram representation. 
Then, a back-end deep learning network is used to classify the spectrogram features into categories of respiratory anomaly cycles or diseases.
Experiments, conducted over the ICBHI benchmark dataset of respiratory sounds, confirm three main contributions towards respiratory-sound analysis.
Firstly, we carry out an extensive exploration of the effect of spectrogram type, spectral-time resolution, overlapped/non-overlapped windows, and data augmentation on final prediction accuracy.
This leads us to propose a novel deep learning system, built on the proposed framework, which outperforms current state-of-the-art methods.
Finally, we apply a Teacher-Student scheme to achieve a trade-off between model performance and model complexity which additionally helps to increase the potential of the proposed framework for building real-time applications.

\indent \textit{Clinical relevance}--- Respiratory disease, lung auscultation, wheezes, crackles, anomaly detection, data augmentation, mixture of experts.
\end{abstract}

\section{Introduction}
\label{intro}

According to the World Health Organization (WHO)~\cite{who}, respiratory illness, which comprises lung cancer, tuberculosis, asthma, chronic obstructive pulmonary disease (COPD), and lower respiratory tract infection (LRTI), accounts for a significant percentage of mortality worldwide. 
Indeed, records indicate that around 10 million people currently have tuberculosis (TB), 65 million have chronic obstructive pulmonary disease (COPD) and 334 million have asthma.
Notably, the WHO estimates that about 1.4, 1.6, and 3 million people die from TB, lung cancer or COPD annually, respectively. 
To deal with respiratory diseases,  early detection is the key factor in enhancing the effectiveness of intervention, including treatment and limiting spread.
During a respiratory examination, lung auscultation (listening to the sounds of breathing through a stethoscope) is an important aspect of respiratory disease diagnosis.
By listening to respiratory sounds during lung auscultation, experts can recognise adventitious sounds (including \textit{Crackles} and \textit{Wheezes}) during the respiratory cycle. These often occur in those who have pulmonary disorders.
If automated methods can be developed to detect such anomalous sounds in future, it may improve the early detection of respiratory disease, or indeed enable screening of a wider group of subjects than could be performed manually.
Research into the automated detection or analysis of respiratory sounds has some precedent~\cite{sound_early_01, sound_early_02, sound_early_03}, but has drawn increasing attention in recent years as robust machine hearing methods have been developed, leveraging on ever more capable deep learning techniques.

Most existing respiratory sound analysis systems tend to rely upon frame-based feature representations such as Mel-Frequency Cepstral Coefficients (MFCC)~\cite{sl_hmm_ano_14_embc03, ic_tree_19_embc04}, borrowed from the Automatic Speech Recognition (ASR) and Speaker Recognition (SR) fields. 
However, Gr{\o}nnesby \textit{et al.}~\cite{sl_svm_ano_17_nw} found that MFCCs did not represent crackles well. 
They thus replaced them with five-dimensional feature vectors, comprising four time domain features (variance, range, and sum of simple moving average (coarse \& fine)), and one frequency domain feature (spectrum mean). 
Meanwhile, Hanna \textit{et al.}~\cite{ic_tree_18_cbmi} firstly extracted spectral information from bark, energy and Mel bands, MFCCs, rhythm features from beat loudness, harmonicity and inharmonicity features, as well as tonal features such as chords strength and tuning frequency. 
Next,  they computed statistical features including standard deviation, variance, minimum and maximum, median, mean, first and second derivative mean, and variance of the raw values. 
This extensive list aimed to maximize the chance of achieving a discriminative feature set.
To further explore audio features, Mendes \textit{et al.}~\cite{sl_lg_ano_16_embc02} went further to propose 35 different types of feature, mainly coming from Music Information Retrieval research.
Inspired by the finding that only some features contributed to the final result, Datta \textit{et al.}~\cite{sl_svm_ano_17_embc01} firstly assessed features such as power spectral density (PSD), FFT and Wavelet spectrogram, MFCCs, and Linear Frequency Cepstral Coefficients (LFCC). 
Next, they applied a Maximal Information Coefficient (MIC)~\cite{mic_tech} to score each feature, selecting only the most influencing, before feeding into a classifier to improve performance and reduce complexity slightly. 
Similarly, Kok \textit{et al.}~\cite{ic_tree_19_embc04} applied the Wilcoxon Sum of Rank test to indicate which feature among MFCC,  Discrete Wavelet Transform (DWT) and a set of time domain features (namely power, mean, variance, skewness and kurtosis of audio signal) mainly affected final classification accuracy.
Image processing techniques were then tried by Sengupta \textit{et al.}~\cite{sl_lbsvm_17_fl}, who applied Local Binary Pattern (LBP) analysis on mel-frequency spectral coefficient (MFSC) to capture texture information from the MFSC spectrogram, thus obtained an LBP spectrogram.
The LBP spectrogram was converted into a histogram presentation before feeding it into a back-end classifier, outperforming previous MFCC-based methods.
In these systems, the time stream of audio feature vectors is classified by a range of traditional machine learning models. These include Logistic Regression ~\cite{sl_lg_ano_16_embc02}, $k$-Nearest Neighbour (KNN)~\cite{sl_svm_ano_17_nw, sl_lbsvm_17_fl}, Hidden Markov Models~\cite{sl_hmm_ano_14_embc03,  sl_hmm_19_sas, ic_hmm_18_sp}, Support Vector Machines~\cite{sl_svm_ano_17_nw, sl_svm_ano_17_embc01, sl_lbsvm_17_fl, ic_pca_svm_18_springer01, ic_svm_18_sp} and decision trees~\cite{ic_tree_19_embc04, sl_svm_ano_17_nw, ic_tree_18_cbmi, ic_baseline}. 

Deep learning techniques have achieved strong and robust detection performance for general sound classification~\cite{ivmCNNsounddet},~\cite{ivmearly_2018}.
Feature extraction in state-of-the-art systems typically involves generating two-dimensional time-frequency spectrograms that are able to capture both fine grained temporal and spectral information as well as present a much wider time context than single frame analysis can using traditional featuers.
While a variety of spectrogram transformations have been utilised, Mel-based methods such as log-Mel spectra~\cite{sl_rnn_19_ieeeopen, ic_cnn_19_ici, ic_cnn_20_ieee_bs} and stacked MFCC features~\cite{sl_rnn_19_ieeeopen, sl_cnn_17_eurasip, ic_cnn_18_bibm, ic_rnn_19_cbms, sl_rnn_18_embc05, ic_rnn_18_ann} are the most popular approaches.
Some researchers combined different types of spectrogram, e.g. short-time Fourier transform (STFT) and Wavelet as proposed by Minami \textit{et al.}~\cite{ic_cnn_19_iccas} or optimized S-Transformations in~\cite{ic_trip_19_ieeeopen}.
Although extracting good quality representative spectrogram input features is very important for the back-end classifier, researchers to date have not explored the settings used in this step deeply -- something we aim to contribute in this paper.

Current deep learning classifiers acting on spectrograms for research into respiratory sound analysis are mainly based on Convolutional Neural Networks (CNN), Recurrent Neural Networks (RNN), or hybrid architectures.
The CNN-based systems span some diverse architectures such as LeNet6~\cite{ic_cnn_18_bibm, sl_cnn_17_eurasip}, VGG5~\cite{ic_cnn_19_ici}, two parallel VGG16s~\cite{ic_cnn_19_iccas}, and ResNet50~\cite{ic_trip_19_ieeeopen}.
Inspired by the fact that respiratory indicator sounds such as \textit{Crackle} and \textit{Wheeze} present certain characteristic temporal sequences, RNN-based networks have been developed to attempt to capture time-frequency structures. For example, Perna and Tagarelli~\cite{ic_rnn_19_cbms} analysed the use of a Long Short-term Memory (LSTM) network for two tasks of classifying anomalous respiratory sounds and classifying respiratory diseases.
By using LSTM and Gated Recurrent Unit (GRU) cells in a RNN-based network, Kochetov \textit{et al.}~\cite{ic_rnn_18_ann} proposed a novel architecture, namely the Noise Masking Recurrent Neural Network, which aimed to distinguish both noise and anomalous respiratory sounds.
Hybrid architectures were proposed in ~\cite{ic_cnn_20_ieee_bs, ic_cnn_19_iccas}. A CNN was first used to map a spectrogram input to a temporal sequence. 
Next, LSTM~\cite{ic_cnn_20_ieee_bs} or GRU~\cite{ic_cnn_19_iccas} layers were used to learn sequence structures before classification by fully-connected layers.

State-of-the-art respiratory sound detection performance comparisons presented in~\cite{ic_rnn_19_cbms, ic_cnn_19_iccas, ic_trip_19_ieeeopen} indicate that deep learning classifiers are robust and effective. 
However, some of the deep learning based models have extremely complicated architectures, limiting their implementation within mobile or wearable real-time devices. 
Clearly, state-of-the-art systems involve ever-increasing model complexity.

A more serious issue with this research field has been the difficulty of comparing between techniques due to the lack of standardised datasets used by authors for evaluation. 
Most publications evaluate over proprietary datasets that are unavailable to others~\cite{sl_lg_ano_16_embc02, sl_svm_ano_17_embc01, sl_hmm_19_sas, sl_rnn_19_ieeeopen, sl_rnn_18_embc05}.

In this paper, we tackle the main issues of respiratory sound analysis in the following way;
\begin{itemize}
\item Firstly we ensure repeatability and ease of comparison by adopting the 2017 Internal Conference on Biomedical Health Informatics (ICBHI)~\cite{ic_dataset} dataset for all experiments.
The ICBHI dataset is one of the largest currently available which includes audio recordings. 
Using this resource, we will comprehensively analyse factors such as different types of spectrogram, overlapped/non-overlapped windowing, patch sizes, and data augmentation to pinpoint their effect on performance.

\item From this analysis, we next propose a deep learning framework to target two related tasks of anomaly sound classification and respiratory disease detection. 
We evaluate over two methods of train/test splitting used in the literature (namely random 5-fold cross validation and 60/40 splitting as per the ICBHI challenge recommendation), and compare against state-of-the-art systems.

\item To aid in the trade-off between performance and complexity, we propose a Student-Teacher scheme.
Specifically, the best deep learning framework used for the task of respiratory disease detection and which requires a large number of trainable parameters, is referred to as the Teacher.
We extract classification information from the Teacher model and distill this information to train another network architecture with fewer trainable parameters, referred to as the Student.
Eventually, we successfully obtain a reduced-size Student network which achieves similar performance as the Teacher.
\end{itemize}

\section{ICBHI dataset and our tasks proposed}
\label{icbhi_tasks}

\subsection{ICBHI dataset}
\label{icbhi}

The 2017 ICBHI dataset~\cite{ic_dataset} provides a large database of labelled respiratory sounds comprising 920 audio recordings with a combined duration of 5.5 hours.
The recording lengths are uneven, ranging from from 10 to 90\,s, and were recorded with a wide range of sampling frequencies from 4\,kHz to 44.1\,kHz. 
In total, the dataset contains recordings from 128 patients, who are identified in terms of being healthy or exhibiting one of the following respiratory diseases or conditions: COPD, Bronchiectasis, Asthma, upper and lower respiratory tract infection, Pneumonia, Bronchiolitis.
These respiratory condition labels are linked to audio recording files.
Within each audio recording, four different types of respiratory cycle are presented -- called \textit{Crackle}, \textit{Wheeze}, \textit{Crackle \& Wheeze}, and \textit{Normal}.
These cycles, labelled by experts, include identified onset and offset times.
The cycles have various recording lengths ranging from 0.2\,s up to 16.2\,s, with the number of cycles being unbalanced (i.e. 1864, 886, 506 and 3642 cycles respectively for \textit{Crackle}, \textit{Wheeze}, \textit{Crackle \& Wheeze}, and \textit{Normal}).

\subsection{Main tasks proposed from ICBHI dataset}
\label{tasks}

Given the ICBHI recordings and metadata, this paper evaluates performance over two main tasks.
\newline \textbf{Task 1}, respiratory anomaly classification, is separated into two sub-tasks.
The first aims to classify between four different cycles (\textit{Crackle}, \textit{Wheeze}, \textit{Crackle \& Wheeze}, and \textit{Normal}).
The second sub-task is to classify the four types of cycle into two groups of \textit{Normal} and \textit{Anomaly} sounds (the latter group consisting of \textit{Crackle}, \textit{Wheeze}, both \textit{Crackle \& Wheeze}). 
For convenience, we will identify these as Task 1-1 and Task 1-2, respectively.
\newline \textbf{Task 2}, respiratory disease prediction, also comprises two sub-tasks.
The first aims to classify audio recordings into three groups of disease conditions: \textit{Healthy}, \textit{Chronic Disease} (i.e. COPD, Bronchiectasis and Asthma) and \textit{Non-Chronic Disease} (i.e. upper and lower respiratory tract infection, Pneumonia, and Bronchiolitis).
The second sub-task is for classification into two groups of \textit{Healthy} or \textit{Unhealthy} (comprising the \textit{Chronic} and \textit{Non-Chronic} disease groups combined).
We name theses sub-tasks Tasks 2-1 and Task 2-2, respectively.
While Tasks 1-1 and 1-2 are evaluated over individual respiratory cycles, Task 2-1 and 2-2 are evaluated over entire audio recordings.

State-of-the-art published systems that use the ICBHI dataset follow two different approaches for splitting the database into training and testing portions.
The first group~\cite{ic_hmm_18_sp, ic_svm_18_sp, ic_baseline, ic_cnn_19_iccas} follow the ICBHI challenge recommendations~\cite{ic_dataset} to divide the dataset into non-overlapping 60\% and 40\% portions for training and test subsets, respectively.
Notably, this avoids a situation in which audio recordings from one subject are found in both of the subsets.
Meanwhile, several other papers~\cite{ic_tree_19_embc04, ic_tree_18_cbmi, ic_cnn_19_ici, ic_cnn_18_bibm, ic_rnn_19_cbms} randomly separate the entire dataset into training and test subsets, with different ratios. 

To evaluate our proposed framework over each task in this paper, we first separate the ICBHI dataset (namely 6898 respiratory cycles for Task 1 and 920 entire recordings for Task 2) into five folds for cross validation.
We next introduce a baseline system upon which we will evaluate the effect of a number of settings and influencing factors. 
Due to extensive training times, this initial exploration evaluates over one fold.
Then, following the initial exploration, we propose two systems; one for the task of anomaly cycle detection (Tasks 1-1 and 1-2) and a second system for respiratory disease detection (Tasks 2-1 and 2-2).
We evaluate each of those systems with both the full 5-fold cross validation and 60/40 splitting as specified in the ICBHI challenge recommendation, and compare against state-of-the-art methods. 

\begin{table}[b]
    \caption{Confusion matrix of anomaly cycle classification.} 
        	\vspace{-0.2cm}
    \centering
    \begin{tabular}{l l l l l} 
        \hline 
                               &  \textbf{Crackle}  &  \textbf{Wheeze} &  \textbf{Both}  &  \textbf{Normal} \\
        \hline 
             \textbf{Crackle}  	& $C_{c}$            & $W_{c}$          & $B_{c}$         & $N_{c}$ \\        
             \textbf{Wheeze}   & $C_{w}$            & $W_{w}$          & $B_{w}$         & $N_{w}$ \\
             \textbf{Both}		& $C_{b}$            & $W_{b}$          & $B_{b}$         & $N_{b}$ \\
             \textbf{Normal}   & $C_{n}$            & $W_{n}$          & $B_{n}$         & $N_{n}$ \\
       \hline 
             \textbf{Total}    	& $C_{t}$            & $W_{t}$          & $B_{t}$         & $N_{t}$ \\
       \hline 
    \end{tabular}
    \label{table:cycle_tab} 
\end{table}
%
\begin{table}[tb]
    \caption{Confusion matrix of respiratory disease detection.} 
        	\vspace{-0.2cm}
    \centering
    \scalebox{1}{
    \begin{tabular}{l l l l} 
        \hline 
              &  \textbf{Chronic}  &  \textbf{Non-chronic} &  \textbf{Healthy}       \\
        \hline 
             \textbf{Chronic}     & $C_{c}$    & $NC_{c}$    & $H_{c}$        \\ 
	     \textbf{Non-chronic} & $C_{nc}$ & $NC_{nc}$ & $H_{nc}$     \\
	     \textbf{Healthy}	  & $C_{h}$     & $NC_{h}$     & $H_{h}$         \\ 
       \hline 
	     \textbf{Total}	  & $C_{t}$     & $NC_{t}$     & $H_{t}$         \\            
       \hline 
    \end{tabular}
    }
    \label{table:disease_tab} 
\end{table}

\subsection{Evaluation metrics}
\label{tasks}

The baseline and proposed framework variants are assessed using the metrics of \textit{Sensitivity} (Sen.), \textit{Specitivity} (Spec.), and \textit{ICBHI score}~\cite{ic_rnn_19_cbms, ic_dataset}.
To understand these scores, consider a confusion matrix for Task 1 as presented in Table \ref{table:cycle_tab}. 
In this case, \textit{C, W, B,} and \textit{N} denote the numbers of cycles of  \textit{Crackle}, \textit{Wheeze}, \textit{Crackle \& Wheeze (Both), and \textit{Normal}, respectively,} whereas \textit{c, w, b,} and \textit{n} subscripts indicate the classification results. The sums $C_{t}$, $W_{t}$, $B_{t}$ and $N_{t}$ are the total numbers of cycles. 

\textit{Sensitivity} is then computed for Task 1-1 (4-class anomaly classification) as follows;
 \begin{equation}
     \label{eq:task_1_1_sen}
     Sensitivity =  \frac{C_{c} + W_{w} + B_{b}}{C_{t} + W_{t} + B_{t}},
 \end{equation}
and for Task 1-2 (binary anomaly classification) as;
 \begin{equation}
     \label{eq:task_1_2_sen}
     Sensitivity =  \frac{C_{c+w+b} + W_{c+w+b} + B_{c+w+b}}{C_{t} + W_{t} + B_{t}}, 
 \end{equation}
where $C_{c+w+b} = C_c + C_w + C_b$, $W_{c+w+b} = W_c + W_w + W_b$, and $B_{c+w+b} = B_c + B_w + B_b$. Then we can define 

 \begin{equation}
Specificity =  \frac{N_{n}}{N_{t}}.
 \end{equation}

Similarly, consider the Task 2 confusion matrix as shown in Table \ref{table:disease_tab}. 
In this case, \textit{C, NC} and \textit{H} are the numbers of recordings of the three Task 2 classes. \textit{c, nc} and \textit{h} subscripts indicate the classification results. As before, $C_{t}$, $NC_{t}$, and $H_{t}$ are the total numbers of \textit{Chronic}, \textit{Non-chronic}, and \textit{Healthy} recordings, respectively.

For Task 2-1,  \textit{Sensitivity} is defined as follows;
\begin{equation}
     \label{eq:task_2_1_sen}
     Sensitivity =  \frac{C_{c} + NC_{nc}}{C_{t} + NC_{t}},
 \end{equation}
and for Task 2-2 it reads;
  \begin{equation}
     \label{eq:task_2_2_sen}
     Sensitivity =  \frac{(C_{c} + C_{nc}) + (NC_{c} + NC_{nc})}{C_{t} + NC_{t}}.
 \end{equation}
We simply then define 
  \begin{equation}
Specificity =  \frac{H_{h}}{H_{t}}.
 \end{equation}
Regarding the \textit{ICBHI score}, this represents an equal trade-off between the two metrics and is computed in the same way for each task -- namely averaging the \textit{Sensitivity} and the \textit{Specificity} scores. \\

\section{High-level framework architecture}
\label{architecture}

\subsection{High-level description}
\label{high_level}
\begin{figure*}[h!]
    \centering
    \includegraphics[width =0.9\linewidth]{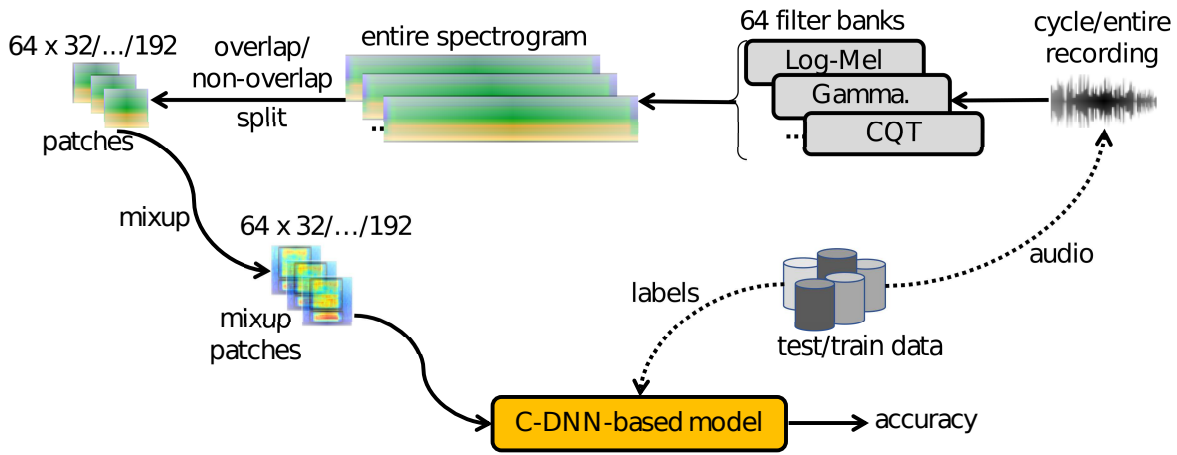}
	\caption{The high-level architecture and processing sequence of the proposed framework.}
    \label{fig:high_level_arc}
\end{figure*}
The proposed high-level system architecture used for all the tasks is described in Fig. \ref{fig:high_level_arc}.
The architecture is divided into two main parts: front-end feature extraction (the upper part) and back-end deep learning models (the lower part).
In general, respiratory cycles in Task 1 or entire audio recording in Tasks 2 are transformed into one or more spectrogram representations.
The spectrograms are then split into equal-sized image patches.
During training, mixup data augmentation~\cite{mixup1, mixup2} is applied to the patches to generate an expanded set of training data that is fed into a deep learning classifier.

\subsection{Baseline system}
\label{baseline}
\begin{table}[b!]
    \caption{Baseline system settings.} 
        	\vspace{-0.2cm}
    \centering
    \scalebox{1}{
    \begin{tabular}{l c} 
        \hline 
            \textbf{Factors}   &  \textbf{Setting}  \\
        \hline 
             Re-sample  & 16kHz \\         
             Cycle duration (only for Task 1) & 5s \\         
             Spectrogram & log-Mel \\         
             Patch splitting & non-overlapped \\         
             Patch size & $64\times64$ \\         
             Data augmentation & None \\         
             Deep learning model & C-DNN based architecture\\
      \hline 
    \end{tabular}
    }
    \label{table:baseline} 
\end{table}
\begin{table}[b!]
    \caption{Baseline C-DNN network architecture} 
        	\vspace{-0.2cm}
    \centering
    \scalebox{0.8}{
    \begin{tabular}{l l c} 
        \hline 
            \textbf{Architecture} & \textbf{layers}   &  \textbf{Output}  \\
        \hline 
             & Input layer (image patch)  &          $64{\times}64$   \\
         Conv. Block 01   & Bn - Cv [$3{\times}3$] - Relu - Bn - Ap [$2{\times}2$] - Dr ($10\%$)      & $32{\times}32{\times}64$\\
         Conv. Block 02  & Bn - Cv [$3{\times}3$] - Relu - Bn - Ap [$2{\times}2$] - Dr ($15\%$)      & $16{\times}16{\times}128$\\
         Conv. Block 03  & Bn - Cv [$3{\times}3$] - Relu - Bn - Dr ($20\%$)      & $16{\times}16{\times}256$ \\
         Conv. Block 04  & Bn - Cv [$3{\times}3$] - Relu - Bn - Ap [$2{\times}2$] - Dr ($20\%$)       & $8{\times}8{\times}256$\\
         Conv. Block 05  & Bn - Cv [$3{\times}3$] - Relu - Bn  - Dr ($25\%$)      & $8{\times}8{\times}512$ \\
         Conv. Block 06  & Bn - Cv [$3{\times}3$] - Relu - Bn -  Gap - Dr ($25\%$) & $512$ \\           
         Dense Block & Fl - Softmax layer & $C$ \\         
       \hline 
    \end{tabular}
    }
    \label{table:C-DNN} 
\end{table}

From the high-level architecture shown in Fig. \ref{fig:high_level_arc}, it can be seen that a variety of factors could affect the performance of the classifier. 
These include the type of spectrogram used, the size of image patches and their degree of overlap, and the application of data augmentation. 
We are thus prompted, in this paper, to investigate the most influencing factors among those listed above. 
To limit the investigation scope to manageable proportions, we constrain the deep learning architecture assessed, thus we propose a \textbf{C-DNN} baseline like VGG-7~\cite{vgg_net}, defined below.

The main characteristics and settings of this baseline architecture are listed in Table \ref{table:baseline}, while the network architecture is presented in Table~\ref{table:C-DNN}.

During processing, we first re-sample all audio recordings (which, as aforementioned, were stored with a variety of sample rates) to 16\,kHz mono.
Since respiratory cycle lengths differ quite widely, we repeat short cycles to ensure that input features for Task 1 have a minimum length of 5\,s or longer. This is of course unnecessary for Task 2 which uses entire recordings.
Next, each cycle (for Task 1) or recording (for Task 2) is transformed into a spectrogram with 64 features per analysis frame.
For example, the log-Mel spectrogram is formed from 1024 sample windows over a 256 sample hop, has an FFT length of 2048 and average pools in the frequency domain to yield 64 output bins.
Whichever type of spectrogram is used, the resulting time-frequency output is split into square non-overlapping patches of dimension $64{\times}64$.

Since data augmentation is one of factors evaluated, we do not apply this technique on the baseline system.
Looking at the network architecture of Table~\ref{table:C-DNN} in more detail, we see seven blocks -- six are convolutional and one is a dense block. The former blocks comprise batch normalization (Bn) layers, convolutional (Cv[kernel size]) layers, rectified linear units (Relu), average  (Ap[kernel size]) and global average pooling (Gap) layers, and use dropout (Dr (dropout percentage)). The dense block comprises a fully-connected (Fl), and a final Softmax layer for classification.
\textit{C} refers to the number of classes, which depend on the specific task being evaluated. 
i.e. we train and test two separate \textbf{C-DNN} models with \textit{C} set to 4 and 3 for Tasks 1 and 2, respectively.

\subsection{Experimental settings for the baseline system}
\label{baseline}

All the systems are implemented using TensorFlow. Network training makes use of the Adam optimiser~\cite{adam_tool} with $100$ training epochs, a mini batch size of $100$, and cross-entropy loss:
\begin{equation}
    \label{eq:entro_loss}
    L_{Entropy}(\theta) = -\frac{1}{N}\sum_{i=1}^{N}\mathbf{y}_i\log \mathbf{\hat{y}}_{i}(\theta) + \frac{\lambda}{2}||\theta||_{2}^{2},
\end{equation}
where \(\theta\) are all trainable parameters, \(N\) is batch size, and constant \(\lambda\) is empirically set to $0.0001$.  
$\mathbf{y}_{i}$ and $\mathbf{\hat{y}}_{i}$ denote expected and predicted results, respectively.

An entire spectrogram or cycle is separated into smaller patches and applied patch-by-patch to the \textbf{C-DNN} model which then returns the posterior probability computed over each patch. 
The posterior probability of an entire spectrogram is the average of all patches' posterior probabilities.
Let us consider $\mathbf{P}^{n} = (P_{1}^{n}, P_{2}^{n},\ldots,P_{C}^{n})$ the posterior probability obtained from the \(n^{th}\) out of \(N\) patches. Then, the mean posterior probability of a test sound instance is denoted as \(\mathbf{\bar{P}} = (\bar{P}_{1}, \bar{P}_{2},\ldots, \bar{{P}}_{C})\) where
\begin{equation}
\label{eq:mean_stratergy_patch}
\bar{P}_{c} = \frac{1}{N}\sum_{n=1}^{N}P_{c}^{n}  ~~~  \text{for}  ~~ 1 \leq n \leq N.
\end{equation}
The predicted label  \(\hat{y}\) is then determined as 
\begin{equation}
\hat{y} = \argmax_{c \in \{1,2,\ldots,C\}}\bar{P}_c.
\end{equation}



\section{Analysis of influencing factors}
\label{analyse}

We conducted experiments using the  baseline system to investigate the  impact of various factors on performance.

\subsection{Influence of spectrogram type}
\label{spec}
From our previous work on natural sound datasets~\cite{lam_dca_16_int, lam_dca_18}, we have established that the choice of spectrogram is one of the most important factors to affect final classification accuracy.
Therefore, we now evaluate the effect of spectrogram type on ICBHI performance for each task.
To this end, we maintain all settings as described in Table \ref{table:baseline} but use four spectrogram types: log-Mel spectrogram~\cite{librosa_tool}, Gammatone filter (Gamma)~\cite{gam_tool} spectrogram, stacked Mel-Frequency Cepstral Coefficients (MFCC)~\cite{librosa_tool}, and rectangular Constant Q Transform (CQT)~\cite{librosa_tool} spectrogram. We will evaluate each of the spectrogram types on all four subtasks.

\begin{figure}[t!]
	\centering
	\includegraphics[width =\linewidth]{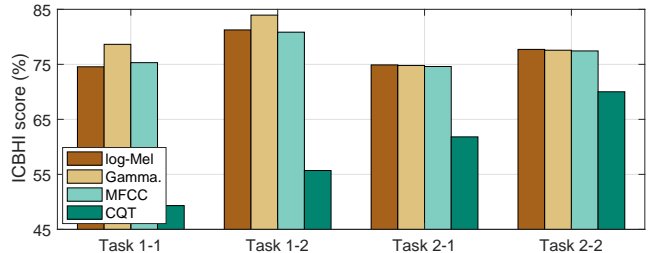}
	\vspace{-0.8 cm}
	\caption{Comparison of baseline performance using different spectrograms.}
	\label{fig:spec_01}
\end{figure}

The ICBHI Score results are plotted in Fig.~\ref{fig:spec_01}, revealing that MFCC, log-Mel, and Gamma spectrogram perform competitively, and are much better than the CQT for all subtasks. 
Compared to log-Mel, the Gamma spectrogram results achieve an improvement in ICBHI score of 4\% for Task 1-1 and 2.7\% for Task 1-2. 
However log-Mel slightly outperforms its Gamma counterpart for Task 2 (0.1\% and 0.2\% for the two subtasks).
MFCC, meanwhile, improves on log-Mel for Task 1-1 (0.8\%) but is worse for all other subtasks (-0.4\%, -0.3\% and -0.3\%, respectively).

These results suggest that the Gamma spectrogram is optimal for anomaly cycle classification (Task 1) while the log-Mel spectrogram works best for detection of respiratory diseases (Task 2). We thus adopt these two spectrograms in the following experiments for those respective tasks.

\subsection{Influence of degree of overlap}
\label{overlap}
\begin{table}[t!]
    \caption{Baseline performance loss or gain on each subtask when overlapping spectrogram patches are used.}
        	\vspace{-0.2cm}
    \centering
    \begin{tabular}{l c c c c} 
        \hline 
           & \textbf{Task 1-1} & \textbf{Task 1-2}   &  \textbf{Task 2-1} &  \textbf{Task 2-2}  \\
        \hline 
         No overlap & \textbf{78.6} &\textbf{84.0}  &74.9                &77.2\\
        Overlap          &  77.8              & 83.7                &\textbf{76.6} & \textbf{78.6} \\
         \hline 
    \end{tabular}
    \label{table:overlap} 
\end{table}
As spectrogram representations of an entire cycle or audio recoding are long in terms of temporal dimension and are of variable length, they are split into smaller patches of $64\times64$ for presentation to the back-end deep learning models. 
In traditional signal processing systems, overlapping analysis windows are used to prevent occlusion of important features in the original data by edge effects. 
We therefore consider the effect of overlapped or non-overlapped patches on ICBHI performance.
Specifically, we compare the baseline with no overlap (the settings in Table \ref{table:baseline}), with a system in which patches are overlapped by 50\% (noting that Gamma and log-Mel are applied on Task 1 and Task 2, respectively).
Results shown in Table \ref{table:overlap} interestingly reveal that Task 1 performs better with non-overlapped patches (subtask scores of 78.6\% and 84.0\%, respectively) while Task 2 performs better with overlapped patches (subtask scores of 76.6\% and 78.6\%, respectively).
We note two contributory factors which may explain this; firstly that both tasks now use different spectrogram types, and secondly that Task 1 classifies respiratory cycles that are repeated in the case of short cycle data, whereas Task 2 classifies unrepeated recordings.

\subsection{Influence of time resolution}
\label{resolution}

The baseline network operates on fixed size patches where the time span encoded in each patch is defined by its horizontal dimension and sampling rate.
Features are presented sequentially, and so the time span also defines the temporal resolution of features presented to the classifier.
In this section, we explore the effect of changing temporal resolution by adjusting patch widths to 0.6\,s, 1.2\,s, 1.8\,s, 2.4\,s, and 3.0\,s. This is achieved  by changing the patch dimension to be $64{\times}32$, $64{\times}64$, $64{\times}96$, $64{\times}128$, and $64{\times}160$, respectively, then training and evaluating the performance of each system on each task.
We note that all settings are reused from Table \ref{table:baseline} with exception that the Gamma and log-Mel spectrograms are used for Task 1 and Task 2, respectively. The frequency resolution (vertical dimension) remains unchanged in each case. The dimension of the network input layer is widened or narrowed to accommodate the differing time resolution.

\begin{figure}[t!]
	\centering
	\includegraphics[width =\linewidth]{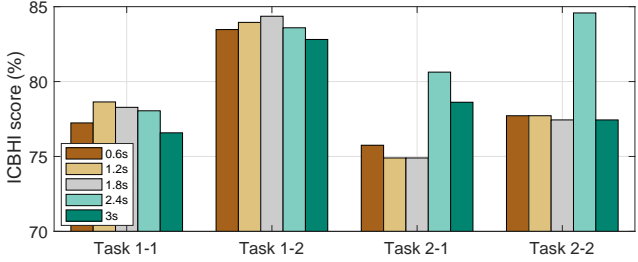}
	\vspace{-0.8 cm}
	\caption{Performance comparison between different time resolutions on each task.}
	\label{fig:time_res}
\end{figure}

Results are shown in Fig. \ref{fig:time_res} for the four subtasks. 
Patches of size $64\times64$ (i.e. 1.2\,s time resolution) as in the baseline system perform best for Task 1-1 and second best for Task 1-2 (scoring 78.6\% and 84.0\%, respectively). 
However we note that a double sized patch, $64\times128$ (i.e. 2.4\,s time resolution) clearly outperforms all other alternatives for Tasks 2-1 and 2-2 (scoring 80.6\% and 84.6\%, respectively).

\subsection{Influence of data augmentation}
\label{augmentation}
\begin{figure*}[th]
    \centering
    \includegraphics[width =0.9\linewidth]{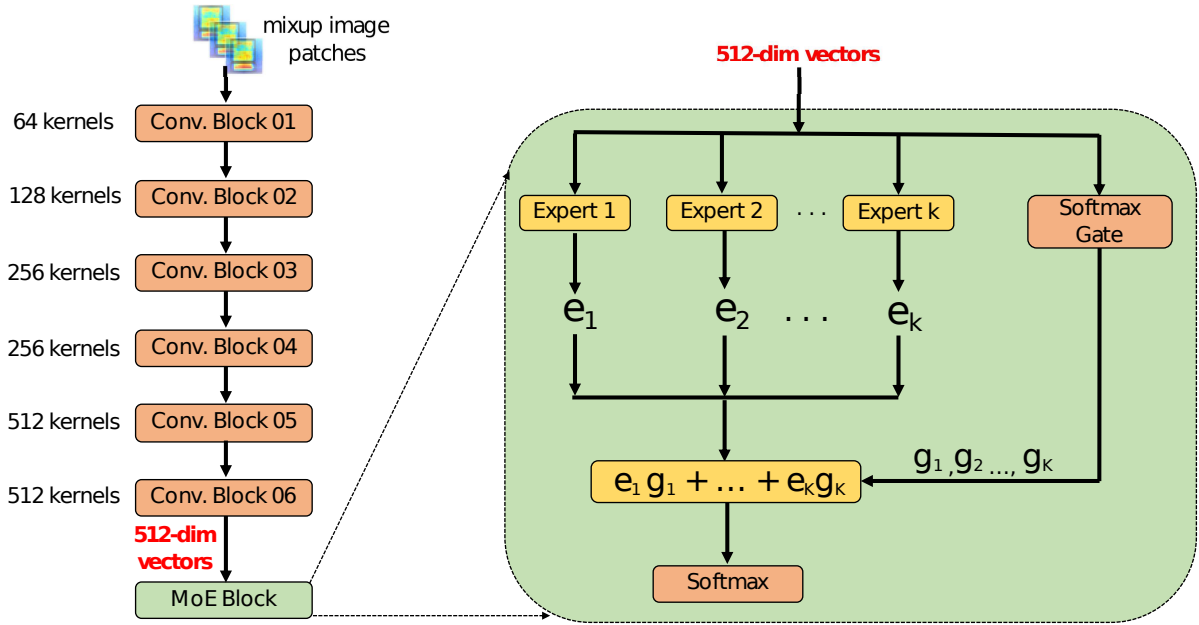}
    	\vspace{-0.2 cm}
    	\caption{The proposed CNN-MoE architecture.}
    \label{fig:framework}
\end{figure*}

Data augmentation (DA) has been shown useful to improve the learning ability of deep learning models in research involving natural sound classification~\cite{lam_dca_18, lam_dca_16_int}.
We, therefore, apply DA in the form of mixup~\cite{mixup1, mixup2} and evaluate its effect on respiratory sound classification.
Consider $\mathbf{X}_1$ and $\mathbf{X}_2$ to be two image patches randomly selected from the original image patches with their corresponding labels $\mathbf{y}_1$ and $\mathbf{y}_2$. Mixup DA  generates new image patches:
\begin{equation}
    \label{eq:mix_up_x1}
    \mathbf{X}_{mp1} = \alpha\mathbf{X}_{1} + (1-\alpha)\mathbf{X}_{2},
\end{equation}
\begin{equation}
    \label{eq:mix_up_x2}
    \mathbf{X}_{mp2} = (1-\alpha)\mathbf{X}_{1} + \alpha\mathbf{X}_{2},
\end{equation}
\begin{equation}
    \label{eq:mix_up_y}
    \mathbf{y}_{mp1} = \alpha\mathbf{y}_{1} + (1-\alpha)\mathbf{y}_{2},
\end{equation}
\begin{equation}
    \label{eq:mix_up_y}
    \mathbf{y}_{mp2} = (1-\alpha)\mathbf{y}_{1} + \alpha\mathbf{y}_{2},
\end{equation}
where 
$\mathbf{X}_{mp1}$ and $\mathbf{X}_{mp2}$ are the two new image patches obtained by mixing $\mathbf{X}_1$ and $\mathbf{X}_2$ with a mixing coefficient $\alpha$.
By using two types of uniform or beta distribution to generate mixing coefficient $\alpha$, this doubles the data size and hence, the training time. Note that for Task 1 the DA mixes the \textit{Normal} class with one of the other classes (since there is already one mixed class in the dataset, i.e. \textit{Crackle \& Wheeze}), whereas it randomly mixes samples of all classes for Task 2.
After mixup, the generated patches are shuffled and fed into the \textbf{C-DNN} baseline. 
Since the labels $\mathbf{y}_{mp1}$ and $\mathbf{y}_{mp2}$ of the resulting patches are no longer one-hot encoded, it is, therefore, necessary to replace the cross-entropy loss by the Kullback-Leibler (KL) divergence loss:
\begin{align}
   \label{eq:kl_loss}
   L_{KL}(\theta) = \sum_{n=1}^{N}\mathbf{y}_{n}\log \left\{ \frac{\mathbf{y}_{n}}{\mathbf{\hat{y}}_{n}} \right\}  +  \frac{\lambda}{2}||\theta||_{2}^{2}.
\end{align}
Again, $\theta$ denotes the trainable network parameters and $\lambda$ denote the $\ell_2$-norm regularization coefficient, set to 0.0001. \(N\) is the batch number,
$\mathbf{y}_{n}$ and $\mathbf{\hat{y}}_{n}$  denote the ground-truth and the network output, respectively.
%
\begin{table}[h]
    \caption{Performance with and without mixup data augmentation.} 
        	\vspace{-0.2cm}
    \centering
    \begin{tabular}{l c c c c} 
        \hline 
           & \textbf{Task 1-1} & \textbf{Task 1-2}   &  \textbf{Task 2-1} &  \textbf{Task 2-2}  \\
        \hline 
         Non-mixup & 78.6              &84.0               &74.9                &77.2\\
         mixup         & \textbf{79.8} &\textbf{84.7} & \textbf{83.5} & \textbf{85.4} \\
         \hline 
    \end{tabular}
    \label{table:augmentation} 
\end{table}

Using the settings in Table \ref{table:baseline} with the Gamma spectrogram for Task 1 and the log-Mel spectrogram for Task 2, we can assess the improvement over the baseline ICBHI score for each subtask due to mixup data augmentation.
Results shown in Table \ref{table:augmentation} indicate that mixup data augmentation substantially improves Task 2 scores (by 8.6\% and 8.2\% on Tasks 2-1 and 2-2, respectively). However, much less improvement is seen for Task 1.

\section{Enhanced deep learning framework}
\label{proposed_framework}
\begin{table}[t!]
    \caption{Deep learning frameworks for Tasks 1 and 2.} 
        	\vspace{-0.2cm}
    \centering
    \begin{tabular}{l c c} 
        \hline 
            \textbf{Factors}   &  \textbf{Anomaly cycle}  &  \textbf{Respiratory disease}  \\
                                         &  \textbf{classification}  &  \textbf{detection} \\
        \hline 
             Resample  & 16kHz & 16kHz\\         
             Cycle duration & 5s & N/A \\         
             Spectrogram & Gamma & log-Mel \\         
             Patch splitting & non-overlapped & overlapped \\         
             Patch size & $64\times64$ & $64\times128$ \\         
             Data augmentation & Yes & Yes \\      
     \hline 
    \end{tabular}
    \label{table:framework} 
\end{table}
From the analysis of influencing factors presented above, we propose two systems. 
One for Task 1 anomaly cycle classification, and the other for Task 2 respiratory disease detection, both summarised in  Table \ref{table:framework}.
In this section we propose to enhance the performance of the C-DNN architecture by incorporating a mixture-of-experts (MoE) technique into the DNN part of the network, leading to a CNN-MoE architecture.

\subsection{CNN-MoE network architecture}
\label{framework_architecture}

Reviewing the \textbf{C-DNN} architecture as listed in Table \ref{table:baseline}, we note that the first six convolutional blocks are used to map the image patch input to condensed and discriminative embeddings, often referred to as high-level features.
That system next uses a dense block comprising a fully-connected layer and a Softmax layer to classify the features.
On the basis that the embedding may contain more information than a single fully-connected layer can unlock, we replace the dense block with a mixture-of-experts (MoE) block as shown in Fig. \ref{fig:framework}.
The MoE block architecture~\cite{moe_tech} comprises multiple experts connected to a gate network which decides which expert is applied to which input region.
In our context, the 512-dimensional embedding from the final global average pooling layer (Gap) is presented simultaneously to all experts.
The output from all experts is then gated before passing the result through a Softmax layer to determine a final classification output.
In our system, each expert comprises a fully-connected layer and a ReLU activation function. 
Each expert input dimension, as we have noted, is 512, and the output dimension from each is the number or categories \textit{C} classified.
The gate network is implemented as a \textbf{Softmax Gate} --  an additional fully-connected layer with Softmax activation function and a gating dimension equal to the number of experts.
Let  $\mathbf{e}_{1}, \mathbf{e}_{2}, \ldots, \mathbf{e}_{K} \in \mathbf{R}^{C}$ be the output vectors from the $K$ experts, and  $g_{1}, g_{2}, \dots , g_{K}$ be the outputs of the gate network where $g_k \in \mathbf{R}$ and $\sum{g_k}=1$.
The predicted output is then found as,
\begin{equation}
    \label{eq:moe}
    \hat{\mathbf{y}} = \softmax \left\{ \sum_{k=1}^{K} g_{k}\mathbf{e}_{k} \right \}.
\end{equation}

\label{fusion}

The proposed systems, as defined in Table \ref{table:framework}, are trained with KL-divergence loss~\cite{kl_loss} (due to the use of mixup data augmentation) and use the same training settings as the previous experiments with 5-fold cross validation.
\label{compare_state}

\subsection{Performance comparison}
\label{framework_architecture}
\begin{table}[t!]
    \caption{ICBHI score comparison between the C-DNN and CNN-MoE frameworks over 5-fold cross validation and ICBHI challenge splitting (highest scores in \textbf{bold}).} 
        	\vspace{-0.2cm}
    \centering

    \begin{tabular}{l   c c c c } 
        \hline            
           & C-DNN &  CNN-MoE & C-DNN &  CNN-MoE \\
               Tasks      & (5-fold) &  (5-fold) & (ICBHI) &  (ICBHI)\\

        \hline 
        1-1, 4-category                         &77.4              &\textbf{78.5}     &43.3 & \textbf{47.0}\\        
	    1-2, 2-category                             &84.1    &  \textbf{84.4}      &53.3  & \textbf{54.1} \\
	    2-1, 3-category                                   &84.7    &\textbf{90.5}    &78.5 &   \textbf{84.0}         \\
	    2-2, 2-category                        &86.2    &\textbf{92.0}  & 78.7 &  \textbf{84.1} \\
       \hline 
    \end{tabular}
    \label{table:comp_moe} 
\end{table}
\textbf{Comparing C-DNN to CNN-MoE}: We evaluate the efficiency of the applied MoE technique (experimentally using $K$=10 experts) compared to the C-DNN system, reporting the performance of both in Table \ref{table:comp_moe} (note: both the systems follow the settings in Table \ref{table:framework}, with the back-end classifier being either C-DNN or CNN-MoE -- there are thus eight systems trained and evaluated, two C-DNNs and two CNN-MoEs for each type of splitting).
The results in Table \ref{table:comp_moe} clearly indicate that the CNN-MoE systems perform best overall. Although we see only marginal gains over the C-DNN for Task 1, for Tasks 2-1 and 2-2, CNN-MoE achieves a much better improvement of 7\% absolute on 5-fold cross validated random splitting and nearly 6\% on ICBHI challenge splitting. 

\textbf{Comparing to state-of-the-art systems:} We next contrast the proposed framework to state-of-the-art systems.
For each task, we evaluate everything twice -- once for the ICBHI challenge train/test split, and once for random splitting (as described in Section~\ref{tasks}).
Considering first the splitting method specified in the ICBHI challenge, Table \ref{table:comp_sta_icb} presents scores obtained by the proposed framework and state-of-the-art published systems (where available). 
The highest scores for each test are presented in bold.
We note that the proposed framework lies second in terms of Task 1-1 evaluation. 
Our results for other subtasks were listed in Table \ref{table:comp_moe}. Only Task 2-2 is found in the literature (for ICBHI splitting), and the score of 84\% achieved by our proposed system comfortably  outperforms the state-of-the-art result of 72\% ~\cite{ic_cnn_20_ieee_bs}. 

\begin{table}[t!]
    \caption{Comparison against state-of-the-art systems with ICBHI challenge splitting (highest scores in \textbf{bold}).} 
        	\vspace{-0.2cm}
    \centering

    \begin{tabular}{l l l c c c} 
        \hline 
	    \textbf{Task}   &\textbf{Method}                        &\textbf{Spec.}   &\textbf{Sen.}   &\textbf{Score}  \\
        \hline 
        1-1, 4-category      &DT~\cite{ic_baseline}                    &0.75             &0.12           &0.43  \\        
        1-1, 4-category      &HMM~\cite{ic_hmm_18_sp}                   &0.38             &\textbf{0.41}           &0.39  \\        
        1-1, 4-category      &SVM~\cite{ic_svm_18_sp}                   &0.78             &0.20           &0.47  \\
        1-1, 4-category     &CNN-RNN~\cite{ic_cnn_19_iccas}          &\textbf{0.81} &0.28 & \textbf{0.54}    \\ 
	    1-1, 4-category      &\textbf{Our system}                                &0.68    &0.26   &0.47    \\
       \hline 
    \end{tabular}
    \label{table:comp_sta_icb} 
\end{table}

Table \ref{table:comp_sta_rand} now compares performance with previously published results that use the random train/test splitting method. Again, the highest performance for each task is presented in bold text. For Tasks 1-1 and 1-2, the proposed framework clearly outperforms other systems quite comprehensively. 
Meanwhile for Task 2-1 and 2-2 the proposed method also outperforms other systems in terms of overall ICBHI score, but not necessarily simultaneously for both subcomponents of specificity or sensitivity.
\begin{table}[t!]
    \caption{Performance comparison between the proposed system and state-of-the-art systems following random splitting (highest scores are highlighted in \textbf{bold}).} 
        	\vspace{-0.2cm}
    \centering
    \scalebox{0.9}{

    \begin{tabular}{l l l c c c c} 
        \hline 
	    \textbf{Task}   &\textbf{Method}                        &\textbf{train/test}      &\textbf{Spec.}   &\textbf{Sen.}   &\textbf{Score}  \\
        \hline 
	    1-1, 4-category      &Boosted DT~\cite{ic_tree_18_cbmi}  &60/40                    &0.78             &0.21            &0.49  \\
	    1-1, 4-category      &CNN~\cite{ic_cnn_18_bibm}            &80/20                    &0.77             &0.45            &0.61  \\
    	    1-1, 4-category      &CNN-RNN~\cite{ic_cnn_20_ieee_bs}  &5 folds      &0.84             &0.49           &0.66  \\
	    1-1, 4-category      &LSTM~\cite{ic_rnn_19_cbms}           &80/20                    &0.85             &0.62            &0.74  \\
	    1-1, 4-category      &\textbf{Our system}                  &5 folds               &\textbf{0.90}    &\textbf{0.68}   &\textbf{0.79}    \\
       \hline 
	    1-2, 2-category      &Boosted DT~\cite{ic_tree_18_cbmi}  &60/40                    &0.78             &0.33            &0.56  \\
	    1-2, 2-category      &LSTM~\cite{ic_rnn_19_cbms}           &80/20                    &-                &-               &0.81  \\
   	    1-2, 2-category      &CNN~\cite{ic_cnn_19_ici}           &75/25                    &-                &-               &0.82  \\
	    1-2, 2-category      &\textbf{Our system}                  &5 folds               &\textbf{0.90}    &\textbf{0.78}   &\textbf{0.84}    \\
       \hline 
       \hline
	    2-1, 3-category      &CNN~\cite{ic_cnn_18_bibm}            &80/20                    &0.76             &0.89             &0.83  \\
	    2-1, 3-category      &LSTM~\cite{ic_rnn_19_cbms}           &80/20                    &0.82             &\textbf{0.98}    &0.90   \\
	    2-1, 3-category      &\textbf{Our system}                  &5 folds               &\textbf{0.86}    &0.95             &\textbf{0.91}      \\
       \hline 
   	    2-2, 2-category      &Boosted DT~\cite{ic_tree_18_cbmi}  &60/40                    &0.85             &0.85            &0.85  \\
	    2-2, 2-category      &CNN~\cite{ic_cnn_18_bibm}            &80/20                    &0.78             &0.97             &0.88  \\
   	    2-2, 2-category      &RUSBoost DT~\cite{ic_tree_19_embc04}            &50/50                    &\textbf{0.93}             &0.86             &0.90  \\
	    2-2, 2-category      &LSTM~\cite{ic_rnn_19_cbms}           &80/20                    &0.82             &\textbf{0.99}    &0.91 \\
	    2-2, 2-category      &\textbf{Our system}                  &5 folds               &0.86    &0.98    &\textbf{0.92} \\
       \hline 
    \end{tabular}
    }
    \label{table:comp_sta_rand} 
\end{table}
%
\begin{figure*}[h!]
    \centering
    \includegraphics[width =0.9\linewidth]{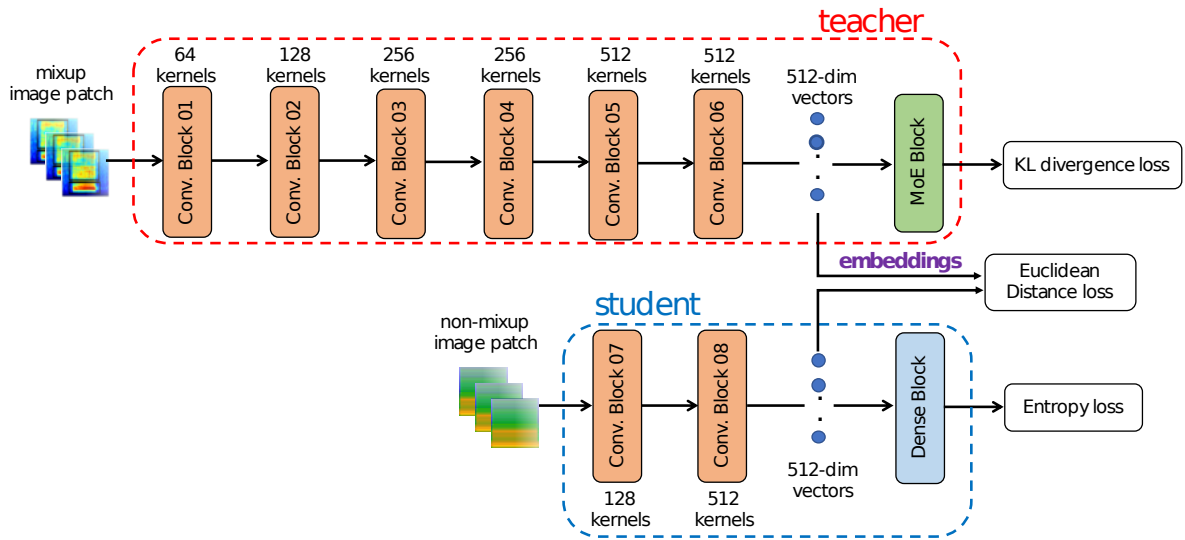}
    \vspace{-0.35cm}
	\caption{Architecture of the Student-Teacher scheme.}
    \label{fig:tea_stu}
\end{figure*}
\subsection{Discussion}
\label{discussion}

Comparing Tables \ref{table:comp_sta_rand} and \ref{table:comp_sta_icb}, it is notable that those systems following the ICBHI splitting recommendation (i.e. recordings from the same patient are never found in both train/test subsets) show considerably lower performance over all tasks than those following random splitting.
This indicates that the ICBHI dataset presents a very high dependence on patient characteristics, which is likely to contribute to the real-world challenge of respiratory cycle classification.

However, all the results obtained from the proposed framework for Tasks 2-1 and 2-2 (with both splitting approaches) exceed 84\%. These results for file-based classification of lung disease -- which is highly related to the overall aim of lung disease detection -- 
provide a strong indicator of the robustness of the underlying framework. As does the fact that the same proposed  framework is capable of performing well for all subtasks.

\section{Student-Teacher scheme for respiratory disease detection}

\subsection{The proposed student-teacher arrangement}
\begin{table}[t!]
    \caption{The Student network architecture.} 
        	\vspace{-0.2cm}
    \centering
    \scalebox{0.9}{

    \begin{tabular}{l l c} 
        \hline 
            \textbf{Architecture} & \textbf{Layers}   &  \textbf{Output}  \\
        \hline 
             & Input layer (image patch)  &          $64{\times}128$   \\
         Conv. Block 07   & Cv [$3{\times}3$] - Relu - Ap [$4{\times}4$]  & $16{\times}32{\times}128$\\
         Conv. Block 08  & Cv [$3{\times}3$] - Relu - Gap       & $512$\\
         Dense Block & Fl - Softmax layer & 3 \\         
       \hline 
    \end{tabular}
    }
    \label{table:student} 
\end{table}
\begin{table*}[t!]
    \caption{Performance comparison between Teacher and Student with and without knowledge distilling.}
        	\vspace{-0.2cm}
    \centering
    \scalebox{1}{

    \begin{tabular}{l l    c c c  c  c c c } 
        \hline 
	        &                 &   \multicolumn{3}{c}{\textbf{Five-fold random split}}  &  \multicolumn{4}{c}{\textbf{ICBHI split}}  \\

	    \textbf{Task}        &                                   &\textbf{Spec}  &\textbf{Sen}  &\textbf{ICBHI Score}  &     &\textbf{Spec}  &\textbf{Sen}  &\textbf{ICBHI Score} \\
        \hline
				 &Teacher                            &0.86	&0.95	&0.91	                           &	 &0.71	&0.98	&0.84                    \\
	    2-1, 3-category      &Student w/o knowledge distill            &0.43	&0.94	&0.68	                           &	 &0.41	&0.97	&0.69                    \\
	                         &\textbf{Student w/ knowledge distill}    &0.86	&0.90	&0.88	                           &	 &0.71	&0.98	&0.84                    \\
       \hline 
       \hline 
				 &Teacher                            &0.86	&0.98	&0.92                              &     &0.71	&0.98	&0.84  \\
	    2-2, 2-category      &Student w/o knowledge distill            &0.43	&0.99	&0.71                              &     &0.41	&0.99	&0.70  \\
	                         &\textbf{Student w/ knowledge distill}    &0.86	&0.96	&0.91                              &     &0.71	&0.98	&0.84  \\
       \hline 
    \end{tabular}
    }
    \label{table:student_res} 
\end{table*}
Recent works on sound scene and sound event detection reported the effectiveness of Teacher-Student learning schemes~\cite{teacher_student_01, teacher_student_02}.
Among other advantages, these schemes offer a trade-off between model size and performance. 
Since the complexity of our best performing MoE framework may be a barrier to future real-time implementation, we explore whether a student-teacher scheme can be used to train a much lower complexity architecture to perform similarly well at the task of respiratory disease detection (Task 2).
The proposed architecture, as shown in Figure \ref{fig:tea_stu}, comprises two networks, namely the Teacher and the Student.
The teacher network re-uses the high-performance CNN-MoE architecture introduced in Section \ref{framework_architecture}. 
On the other hand, the student network features a compact architecture, comprising two convolutional blocks (identified \textit{Conv. Blocks 07 and 08} in the figure), and a dense block whose configuration is the same as the one in Table \ref{table:student} (note that the student network does not apply batch normalisation, dropout or mixup data augmentation).

Training the Teacher-Student network is separated into two phases.
First, the Teacher is trained as usual. 
Afterwards, the Teacher's embedding is distilled to the Student's embedding to assist in the Student's learning process. 
We will also empirically investigate the influence of this knowledge distilling on the student network's performance. 
With the presence of this knowledge distilling, training the student network, therefore, aims to minimize two losses: (1) the Euclidean distance between the teacher and student embedding, and (2) the standard cross-entropy loss on the student's output classification. 
The combined loss function is therefore,
%
\begin{equation}
    \label{eq:final_loss}
    L(\theta) = L_{Entropy}(\theta) + \gamma L_{Euclidean}(\theta),
\end{equation}
Here, 
the hyperparameter \(\gamma\) is empirically set to 0.5 to balance the two constituent losses. \(\theta\) represents the trainable parameters of the student network. Other hyper-parameters and settings are inherited from Section \ref{fusion}.

\subsection{Experimental results using the Teacher-Student scheme}

The experimental results obtained by the student network in comparison with the teacher network are shown in Table \ref{table:student_res}.
On the one hand, it can be seen that without knowledge distilling from the teacher network, the small-footprint student network obtains a substantially reduced specificity score, although it maintains sensitivity quite well. This observation is consistent with the overall ICBHI score and can be explained by the simplicity of the network, and its resulting low learning capacity.


On the other hand, distilling knowledge from the teacher significantly boosts the student performance,  yielding specificity, sensitivity, and ICBHI scores that are very competitive to those of the teacher network -- even thought the student network is much smaller and simpler. Regarding model footprint, the student network has $0.6\times 10^6$ parameters, which is approximately one-seventh of the Teacher's $4.5\times 10^6$ parameters. The fact that similar results can be achieved by a run-time system of such simplicity is remarkable.


\section{Conclusion}
This paper has presented a robust deep learning framework for the analysis of respiratory anomalies and detection of lung disease from lung auscultation recordings.
Extensive experiments were conducted with different architectures and system settings using the ICBHI dataset, and two defined tasks related to that.
The resulting system was been evaluated against existing state-of-the-art methods, outperforming them for most of the challenge tasks.
Furthermore, since the resulting system complexity was such that it may prove a barrier to eventual real-time implementation, a Teacher-Student learning scheme was employed to significantly reduce model complexity while still achieving very high accuracy.
The final experimental results validate the application of deep learning for the timely diagnosis of respiratory disease, bringing this research area one step closer to clinical application.
In future, we aim to explore model compression with pruning and quantisation to further try and reduce complexity, before implementing the simplified detector in an embedded device.

\bibliographystyle{IEEEbib}
\bibliography{refs}

\begin{thebibliography}{10}

\bibitem{who}
World~Health Organization,
\newblock ``The global impact of respiratory diseases (second edition),'' 2017.

\bibitem{sound_early_01}
H{\"u}seyin Polat and {\.I}nan G{\"u}ler,
\newblock ``A simple computer-based measurement and analysis system of
  pulmonary auscultation sounds,''
\newblock {\em Journal of medical systems}, vol. 28, no. 6, pp. 665--672, 2004.

\bibitem{sound_early_02}
R.~J. {Riella}, P.~{Nohama}, R.~F. {Borges}, and A.~L. {Stelle},
\newblock ``Automatic wheezing recognition in recorded lung sounds,''
\newblock in {\em Proc. EMBC}, Sep. 2003, vol.~3, pp. 2535--2538.

\bibitem{sound_early_03}
Sandra Reichert, Raymond Gass, Christian Brandt, and Emmanuel Andr{\`e}s,
\newblock ``Analysis of respiratory sounds: state of the art,''
\newblock in {\em Proc. CCRPM}, 2008, pp. CCRPM--S530.

\bibitem{sl_hmm_ano_14_embc03}
T.~Okubo, N.~Nakamura, M.~Yamashita, and S.~Matsunaga,
\newblock ``Classification of healthy subjects and patients with pulmonary
  emphysema using continuous respiratory sounds,''
\newblock in {\em Proc. EMBC}, 2014, pp. 70--73.

\bibitem{ic_tree_19_embc04}
Xuen~Hoong Kok, Syed~Anas Imtiaz, and Esther Rodriguez-Villegas,
\newblock ``A novel method for automatic identification of respiratory disease
  from acoustic recordings,''
\newblock in {\em Proc. EMBC}, 2019, pp. 2589--2592.

\bibitem{sl_svm_ano_17_nw}
M.~Gr{\o}nnesby, J.~Solis, E.~Holsb{\o}, H.~Melbye, and L.~Bongo,
\newblock ``Feature extraction for machine learning based crackle detection in
  lung sounds from a health survey,''
\newblock {\em arXiv preprint arXiv:1706.00005}, 2017.

\bibitem{ic_tree_18_cbmi}
Ga{\"e}tan Chambres, Pierre Hanna, and Myriam Desainte-Catherine,
\newblock ``Automatic detection of patient with respiratory diseases using lung
  sound analysis,''
\newblock in {\em Proc. CBMI}, 2018, pp. 1--6.

\bibitem{sl_lg_ano_16_embc02}
L.~Mendes, I.~M. Vogiatzis, E.~Perantoni, E.~Kaimakamis, I.~Chouvarda,
  N.~Maglaveras, J.~Henriques, P.~Carvalho, and R.~P. Paiva,
\newblock ``Detection of crackle events using a multi-feature approach,''
\newblock in {\em Proc. EMBC}, 2016, pp. 3679--3683.

\bibitem{sl_svm_ano_17_embc01}
S.~Datta, A.~D. Choudhury, P.~Deshpande, S.~Bhattacharya, and A.~Pal,
\newblock ``Automated lung sound analysis for detecting pulmonary
  abnormalities,''
\newblock in {\em Proc. EMBC}, 2017, pp. 4594--4598.

\bibitem{mic_tech}
D.~Reshef, Y.~Reshef, H.~Finucane, S.~Grossman, G.~McVean, P.~Turnbaugh,
  E.~Lander, M.~Mitzenmacher, and P.~Sabeti,
\newblock ``Detecting novel associations in large data sets,''
\newblock {\em science}, vol. 334, no. 6062, pp. 1518--1524, 2011.

\bibitem{sl_lbsvm_17_fl}
Nandini Sengupta, Md~Sahidullah, and Goutam Saha,
\newblock ``Lung sound classification using local binary pattern,''
\newblock {\em arXiv preprint arXiv:1710.01703}, 2017.

\bibitem{sl_hmm_19_sas}
Dinko Oletic, Marko Matijascic, Vedran Bilas, and Michele Magno,
\newblock ``Hidden markov model-based asthmatic wheeze recognition algorithm
  leveraging the parallel ultra-low-power processor (pulp),''
\newblock in {\em 2019 IEEE Sensors Applications Symposium}, 2019, pp. 1--6.

\bibitem{ic_hmm_18_sp}
Nik{\v{s}}a Jakovljevi{\'c} and Tatjana Lon{\v{c}}ar-Turukalo,
\newblock ``Hidden markov model based respiratory sound classification,''
\newblock in {\em Precision Medicine Powered by pHealth and Connected Health},
  pp. 39--43. Springer, 2018.

\bibitem{ic_pca_svm_18_springer01}
Gorkem Serbes, Sezer Ulukaya, and Yasemin~P Kahya,
\newblock ``An automated lung sound preprocessing and classification system
  based onspectral analysis methods,''
\newblock in {\em Precision Medicine Powered by pHealth and Connected Health},
  2018, pp. 45--49.

\bibitem{ic_svm_18_sp}
Gorkem Serbes, Sezer Ulukaya, and Yasemin~P Kahya,
\newblock ``An automated lung sound preprocessing and classification system
  based on spectral analysis methods,''
\newblock in {\em Precision Medicine Powered by pHealth and Connected Health},
  pp. 45--49. Springer, 2018.

\bibitem{ic_baseline}
B.~M. Rocha, D.~Filos, L.~Mendes, G.~Serbes, S.~Ulukaya, Y.~P. Kahya,
  N.~Jakovljevic, T.~L. Turukalo, I.~M. Vogiatzis, E.~Perantoni, et~al.,
\newblock ``An open access database for the evaluation of respiratory sound
  classification algorithms,''
\newblock {\em Physiological measurement}, vol. 40, no. 3, pp. 035001, 2019.

\bibitem{ivmCNNsounddet}
Haomin Zhang, Ian McLoughlin, and Yan Song,
\newblock ``Robust sound event recognition using convolutional neural
  networks,''
\newblock in {\em Proc. ICASSP}, Apr. 2015, number 2635, pp. 559--563.

\bibitem{ivmearly_2018}
Ian McLoughlin, Yan Song, Lam~Dam Pham, Huy Pham, Palaniappan Ramaswamy, and
  Lang Yue,
\newblock ``Early detection of continuous and partial audio events using
  {CNN},''
\newblock in {\em Proc. INTERSPEECH}, 2018.

\bibitem{sl_rnn_19_ieeeopen}
Lukui Shi, Kang Du, Chaozong Zhang, Hongqi Ma, and Wenjie Yan,
\newblock ``Lung sound recognition algorithm based on vggish-bigru,''
\newblock {\em IEEE Access}, vol. 7, pp. 139438--139449, 2019.

\bibitem{ic_cnn_19_ici}
Renyu Liu, Shengsheng Cai, Kexin Zhang, and Nan Hu,
\newblock ``Detection of adventitious respiratory sounds based on convolutional
  neural network,''
\newblock in {\em Proc. ICIIBMS}, 2019, pp. 298--303.

\bibitem{ic_cnn_20_ieee_bs}
J~Acharya and A~Basu,
\newblock ``Deep neural network for respiratory sound classification in
  wearable devices enabled by patient specific model tuning.,''
\newblock {\em IEEE transactions on biomedical circuits and systems}, 2020.

\bibitem{sl_cnn_17_eurasip}
Murat Aykanat, {\"O}zkan K{\i}l{\i}{\c{c}}, Bahar Kurt, and Sevgi Saryal,
\newblock ``Classification of lung sounds using convolutional neural
  networks,''
\newblock {\em EURASIP Journal on Image and Video Processing}, vol. 2017, no.
  1, pp. 65, 2017.

\bibitem{ic_cnn_18_bibm}
Diego Perna,
\newblock ``Convolutional neural networks learning from respiratory data,''
\newblock in {\em 2018 IEEE International Conference on Bioinformatics and
  Biomedicine (BIBM)}, 2018, pp. 2109--2113.

\bibitem{ic_rnn_19_cbms}
Diego Perna and Andrea Tagarelli,
\newblock ``Deep auscultation: Predicting respiratory anomalies and diseases
  via recurrent neural networks,''
\newblock in {\em Proc. CBMS}, 2019, pp. 50--55.

\bibitem{sl_rnn_18_embc05}
E.~Messner, M.~Fediuk, P.~Swatek, S.~Scheidl, F.~Smolle-Juttner, H.~Olschewski,
  and F.~Pernkopf,
\newblock ``Crackle and breathing phase detection in lung sounds with deep
  bidirectional gated recurrent neural networks,''
\newblock in {\em Proc. EMBC}, 2018, pp. 356--359.

\bibitem{ic_rnn_18_ann}
K.~Kochetov, E.~Putin, M.~Balashov, A.~Filchenkov, and A.~Shalyto,
\newblock ``Noise masking recurrent neural network for respiratory sound
  classification,''
\newblock in {\em International Conference on Artificial Neural Networks},
  2018, pp. 208--217.

\bibitem{ic_cnn_19_iccas}
K.~{Minami}, H.~{Lu}, H.~{Kim}, S.~{Mabu}, Y.~{Hirano}, and S.~{Kido},
\newblock ``Automatic classification of large-scale respiratory sound dataset
  based on convolutional neural network,''
\newblock in {\em Proc. ICCAS}, 2019, pp. 804--807.

\bibitem{ic_trip_19_ieeeopen}
H.~Chen, X.~Yuan, Z.~Pei, M.~Li, and J.~Li,
\newblock ``Triple-classification of respiratory sounds using optimized
  s-transform and deep residual networks,''
\newblock {\em IEEE Access}, vol. 7, pp. 32845--32852, 2019.

\bibitem{ic_dataset}
BM~Rocha, D~Filos, L~Mendes, Vogiatzis, et~al.,
\newblock ``A respiratory sound database for the development of automated
  classification,''
\newblock in {\em Precision Medicine Powered by pHealth and Connected Health},
  pp. 33--37. 2018.

\bibitem{mixup1}
K.~Xu, D.~Feng, H.~Mi, B.~Zhu, D.~Wang, L.~Zhang, H.~Cai, and S.~Liu,
\newblock ``Mixup-based acoustic scene classification using multi-channel
  convolutional neural network,''
\newblock in {\em Pacific Rim Conference on Multimedia}, 2018, pp. 14--23.

\bibitem{mixup2}
Yuji Tokozume, Yoshitaka Ushiku, and Tatsuya Harada,
\newblock ``Learning from between-class examples for deep sound recognition,''
\newblock {\em arXiv preprint arXiv:1711.10282}, 2017.

\bibitem{vgg_net}
Karen Simonyan and Andrew Zisserman,
\newblock ``Very deep convolutional networks for large-scale image
  recognition,''
\newblock {\em arXiv preprint arXiv:1409.1556}, 2014.

\bibitem{adam_tool}
Diederik~P Kingma and Jimmy Ba,
\newblock ``Adam: A method for stochastic optimization,''
\newblock {\em arXiv preprint arXiv:1412.6980}, 2014.

\bibitem{lam_dca_16_int}
Lam Pham, Ian Mcloughlin, Huy Phan, and Ramaswamy Palaniappan,
\newblock ``A robust framework for acoustic scene classification,''
\newblock in {\em Proc. INTERSPEECH}, 09 2019, pp. 3634--3638.

\bibitem{lam_dca_18}
Lam Pham, Ian McLoughlin, Huy Phan, Ramaswamy Palaniappan, and Yue Lang,
\newblock ``Bag-of-features models based on {C-DNN} network for acoustic scene
  classification,''
\newblock in {\em Proc. AES}, 2019.

\bibitem{librosa_tool}
McFee, Brian, R.~Colin, L.~Dawen, Daniel P., M.~Matt, B.~Eric, and N.~Oriol,
\newblock ``librosa: Audio and music signal analysis in python,''
\newblock in {\em Proc. 14th Python in Science Conference}, 2015, pp. 18--25.

\bibitem{gam_tool}
D~P W~(2009) Ellis,
\newblock ``Gammatone-like spectrogram,'' 2009.

\bibitem{moe_tech}
Ekaterina Garmash and Christof Monz,
\newblock ``Ensemble learning for multi-source neural machine translation,''
\newblock in {\em Proc. COLING}, 2016, pp. 1409--1418.

\bibitem{kl_loss}
Solomon Kullback and Richard~A Leibler,
\newblock ``On information and sufficiency,''
\newblock {\em The annals of mathematical statistics}, vol. 22, no. 1, pp.
  79--86, 1951.

\bibitem{teacher_student_01}
Hee-Soo Heo, Jee-weon Jung, Hye-jin Shim, and Ha-Jin Yu,
\newblock ``Acoustic scene classification using teacher-student learning with
  soft-labels,''
\newblock {\em arXiv preprint arXiv:1904.10135}, 2019.

\bibitem{teacher_student_02}
Liang Gao, Haibo Mi, Boqing Zhu, Dawei Feng, Yicong Li, and Yuxing Peng,
\newblock ``An adversarial feature distillation method for audio
  classification,''
\newblock {\em IEEE Access}, vol. 7, pp. 105319--105330, 2019.

\end{thebibliography}


\end{document}